\documentclass[conference]{IEEEtran}

\usepackage{amsmath}
\usepackage{array}
\usepackage{booktabs}
\usepackage{cite}
\usepackage{graphicx}
\usepackage{tabularx}
\usepackage{url}

\newcolumntype{Y}{>{\raggedright\arraybackslash}X}

\title{Exploring the Social Life of Data: Finding Data You Can Trust}

\author{\IEEEauthorblockN{Penny Atkins, Manish Parashar}
\IEEEauthorblockA{\textit{University of Utah}\\
Salt Lake City, Utah, USA}}

\begin{document}

\maketitle

\begin{abstract}
Artificial intelligence is changing the scale and tempo of scientific inquiry. Models can now search, integrate, and reason over data far beyond data repositories familiar to any individual researcher. Yet this expansion creates a prior problem: before a model can produce a trustworthy scientific result, it must locate data that are appropriate for the question, sufficiently reliable for the intended analysis, and accompanied by enough context to support responsible interpretation. As data becomes increasingly abundant, the challenge of finding data has been overcome by the challenge of finding data that you can trust. 

This paper explores how the social and empirical evidence that accumulates when data are used in research can be used, analogous to social trust networks, to determine fit for purpose and trust. Specifically, the paper explores data-usage graphs as a new layer of scientific data infrastructure. A data-usage graph connects datasets to the publications, people, institutions, topics, software, models, workflows, and other datasets through which they are produced and used. These connections reveal the {\it social life of data:} who has relied on a source, for which questions, in what combinations, with which methods, and with what observable impact. They can turn scattered traces of practice into data-usage descriptors that complement conventional metadata and support judgments of trust and fitness for purpose. The central claim is not that popularity establishes trust, but that this can be grown with appropriate contextual history. Usage evidence must therefore be combined with production quality, provenance, governance, semantic clarity, and community validation. The feasibility and value of data usage graphs is demonstrated by implementing the prototype data insights discovery service within the National Data Platform (NDP).
\end{abstract}


\section{Introduction}

\subsection{Trustworthy AI-Enabled Science Needs the Right Data}

The current data renaissance, accelerated by advances in artificial intelligence (AI), is transforming the nature and scale of scientific research across disciplines and becoming a dominant engine for discoveries and innovation. This renaissance is defined not only by the volume and variety of available data, but also by our growing ability to derive knowledge from it using AI-based approaches to find patterns at scale across heterogeneous sources, modalities, and disciplines. However, as data becomes increasingly abundant, the challenge is no longer simply finding data but finding data that can be trusted and is fit for purpose. 
A researcher focused within a mature specialty that is data-rich may know the canonical datasets, the people who created them, the transformations required to use them, their quirks, and the questions they should and should not be asked to answer. However, only a small number of science disciplines have well-curated datasets, and there is a growing volume of data science data~\cite{heidorn2008shedding} that can benefit science workflows with appropriate provenance and governance. Furthermore, as research broadens to cross disciplines and multidisciplinary science workflows integrate diverse datasets, finding trusted data that is fit for purpose becomes significant.

This problem is further exacerbated as we move to agentic workflows, where workflows are composed and orchestrated by AI agents. An AI system searching across domains does not begin with the tacit knowledge that a human researcher may have. Without this knowledge, an agent can retrieve a plausible source whose population, version, granularity, measurement process, licensing, or temporal coverage makes it inappropriate for the task. It can also combine individually credible sources in a way that creates a new, unexamined mismatch. Fluency at the model layer can conceal fragility at the data layer.

For this reason, access to more data does not automatically produce better science. Trustworthy AI-enabled science requires the right data at the right quality level, at the right time, for the right purpose. This is the practical meaning of \emph{fitness for purpose}. It rejects the idea that every scientific task requires perfect data, while also rejecting the idea that technically accessible data are therefore adequate. Fitness for purpose is a relation among a digital object, a scientific question, a proposed operation, a community's standards, and the consequences of error.

The challenge is increasingly urgent because AI systems mediate discovery as well as analysis. When a person or an agent asks, ``Which data should I use to answer this question?'', the system must do more than return results ranked by lexical similarity. It should surface the evidence needed to judge whether a source is \emph{usable}, \emph{useful}, and \emph{used}. Usable data can be accessed and processed under appropriate governance. Useful data align with the intended population, construct, resolution, and task. Used data carry evidence of prior application, feedback, and impact. Together these dimensions, along with provenance and governance, provide a more grounded basis for trust than any one of them alone.

\subsection{Why Finding Trusted Data Is Hard}

The scientific data landscape is fragmented and dynamic. Data are distributed across disciplinary repositories, government portals, institutional stores, project websites, clouds, instruments, and individual research environments. Metadata quality varies widely. Persistent identifiers and standard data citations remain unevenly adopted, while the same dataset may appear under acronyms, aliases, translated names, version labels, URLs, or informal descriptions. Derived data may be subsetted, joined, cleaned, recoded, or forked without a persistent record of the transformation \cite{lane2020,Sostek2024Discovering}.

These conditions create ``dark data'': data that exist and may already be scientifically valuable, but are difficult to discover and contextualize~\cite{heidorn2008shedding}. Catalogs illuminate what has been formally registered, while much scientific practice remains outside that light. Publications, code repositories, notebooks, workflows, data-management records, and community knowledge contain evidence about actual use, but this evidence is scattered and rarely represented as searchable metadata.

Data churn compounds the problem. New sources emerge, existing sources change, and some are retired or disappear. When a widely used dataset is discontinued, it is difficult to determine which research communities depend on it, what downstream findings may be affected, or which substitutes have already been tested. Conversely, a new dataset may have high potential but remain underused because prospective users cannot find examples, tools, or peers that lower their cost of adoption.

The problem is also epistemic. Data quality is multidimensional \cite{wangStrong1996}, and no dataset is trustworthy in the abstract. Accuracy, coverage, reliability, timeliness, objectivity, protection, representativeness, and documentation are both considered and weighted differently for different questions. A high-quality national survey may be unsuitable for estimates at a small geographic scale. Administrative wage records can be strong evidence for quarterly employment and earnings dynamics but weak evidence for the skill content of jobs when occupation, hours, and education are absent. Applying a single global quality label to a dataset would obscure these distinctions.

Finally, the relevant evidence is socially distributed. Producers know how data were generated. Stewards know lineage, validation rules, and access constraints. Domain scientists know common interpretations and failure modes. Data engineers know transformations and operational dependencies. Reusers know whether a source worked for a particular task and which complementary assets were required. No single actor holds the full story of data trust.

\section{Finding Trusted Data That Are Fit for Purpose}

\subsection{Trust as a Property of Relationships}

Social science offers a useful, though necessarily imperfect, analogy. Trust in society is rarely determined by a single attribute of an isolated person. Rather, it develops through networks of relationships, histories of interaction, shared meanings, and mechanisms of accountability \cite{mayer1995}. Three broad dimensions are particularly helpful: structural, relational, and cognitive.

Structural evidence concerns the configuration of connections. Strong ties may reflect frequent interaction; network closure can create accountability; central nodes may coordinate or broker activity; and weak ties can bridge otherwise separated communities \cite{coleman1988,granovetter1973}. Relational evidence concerns history: reciprocity, consistency, reputation, and responses to failure. Cognitive evidence concerns shared language, norms, and interpretive frames that make another actor's behavior intelligible.

The analogy clarifies why static data attributes are necessary but insufficient. A checksum can establish that a file has not changed. A provenance record can show where it came from. Documentation can define variables. None of these alone establishes that the dataset will support a specific scientific inquiry. That judgment benefits from evidence about the dataset's position in a community of practice, its record across uses, and the semantic conventions that make those uses interpretable.

The analogy must not be taken too far. A heavily used dataset can be wrong; a new dataset can be excellent; a central dataset can reflect path dependence, prestige, or unequal access; and a dense community can reproduce shared blind spots. Network evidence should therefore be treated as a fallible signal, not a vote on truth. Its value lies in making the basis of a judgment visible and challengeable. The same concern applies to reputation and centrality algorithms: graph position can inform assessment but cannot substitute for evidence of purpose fit \cite{kamvar2003,page1999}.

\begin{table*}[!t]
\caption{Social Trust Dimensions and Data-Usage Graph Analogs}
\label{tab:social-trust}
\centering
\footnotesize
\begin{tabularx}{\textwidth}{@{}p{0.12\textwidth}YYY@{}}
\toprule
\textbf{Dimension} & \textbf{Social-network signal} & \textbf{Data-usage graph signal} & \textbf{Interpretive caution} \\
\midrule
Structural & Topology, closure, centrality, bridges & Substantive-use diversity, co-use networks, cross-domain bridges & Centrality may reflect legacy, prestige, or unequal access \\
Relational & History, reciprocity, track record & Longitudinal reuse, corrections, feedback, version transitions & Publication traces omit many operational uses and failures \\
Cognitive & Shared language, norms, goals & Semantic alignment, documentation, schemas, purpose match & Consensus can conceal shared assumptions or blind spots \\
\bottomrule
\end{tabularx}
\end{table*}

\subsection{The Social Life of Data}

Data acquire meaning as they move. A dataset is collected by an agent for a purpose; documented and governed by a steward; transformed by software; combined with other sources; interpreted through disciplinary concepts; used in publications and decisions; and sometimes corrected through feedback. Each step leaves traces. Taken together, these traces form the social life of the data.

A data-usage graph makes that life computationally accessible. Its nodes can include datasets, dataset versions, variables, publications, projects, researchers, institutions, software packages, models, workflows, instruments, repositories, and scientific questions. Its edges can distinguish \emph{mentioned-in}, \emph{used-in}, \emph{derived-from}, \emph{joined-with}, \emph{processed-by}, \emph{produced-by}, \emph{validated-by}, \emph{questioned-by}, and \emph{superseded-by} relations. Edge attributes can record time, confidence, evidence excerpts, task, domain, version, and provenance.

\begin{figure*}[!t]
\centering
\includegraphics[width=0.92\textwidth]{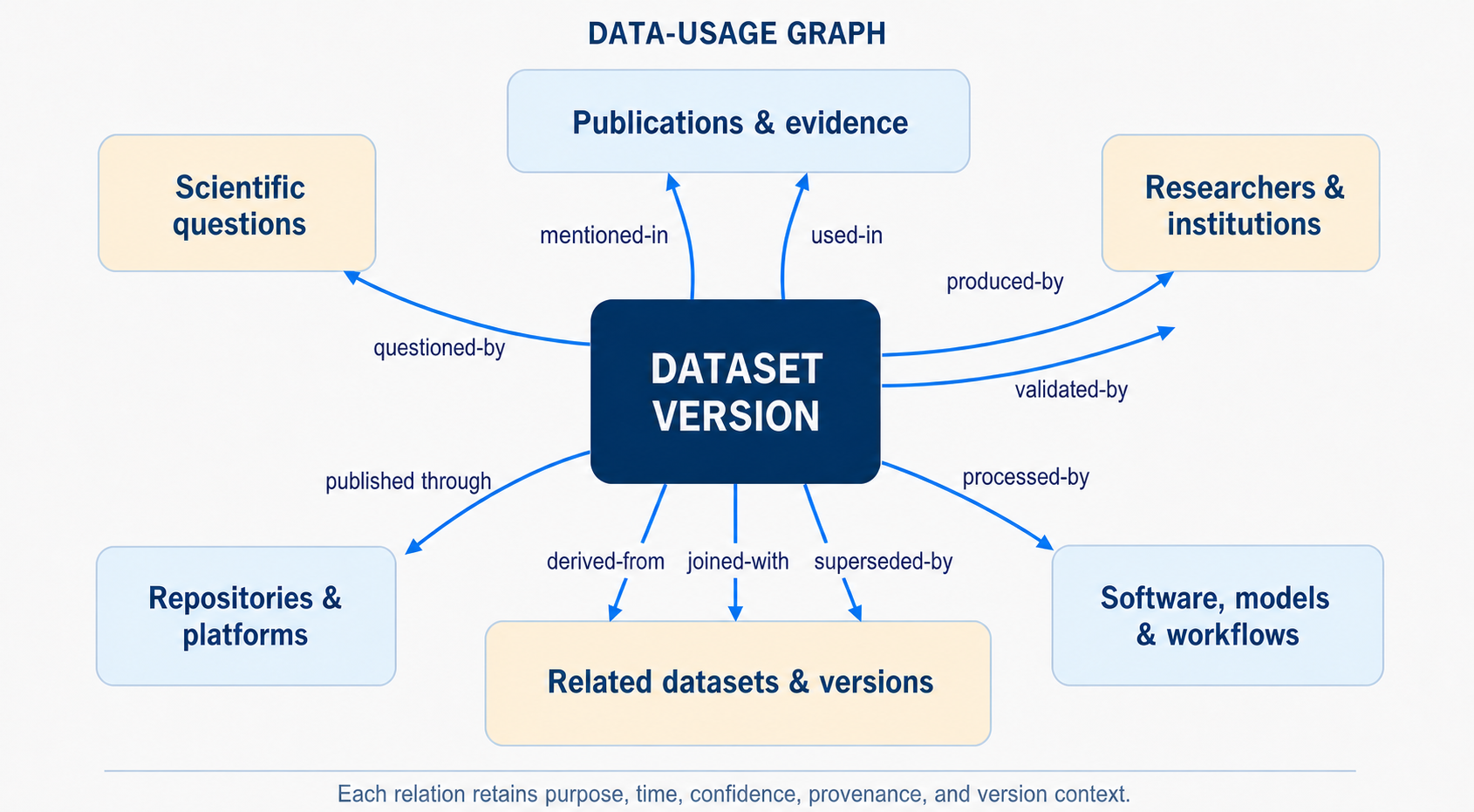}
\caption{A data-usage graph connects datasets and versions to publications, people, institutions, questions, software, models, workflows, and other datasets. Typed, provenance-bearing edges expose the evidence accumulated through use.}
\label{fig:usage-graph}
\end{figure*}

This graph goes beyond counts. It can show that a source has been used by several independent teams to study the same construct; that two datasets are frequently joined; that a particular software package is the dominant route to processing a file format; that a dataset's use is concentrated in one institution; or that a source once considered canonical is declining as a successor emerges. It can also identify people and communities with relevant experience, turning data discovery into a pathway to expertise.

The graph supports several forms of evidence. Structural evidence includes the number and diversity of substantive uses, the dataset's position in co-use networks, and the independence of the communities relying on it. Relational evidence includes longitudinal patterns of successful use, corrections, feedback, version transitions, and reproducible reuse. Cognitive evidence includes agreement in definitions, schema alignment, shared vocabularies, documentation quality, and consistency between a proposed use and prior uses. Production and governance evidence includes provenance, validation controls, stewardship, licensing, security, and access rules.

\subsection{From Usage Evidence to Data Trust}

Data trust can be viewed as a warranted willingness to rely on data for a specified purpose under stated conditions. A useful trust assessment therefore begins with a query of these conditions, not a score: What is the scientific question? Which population, period, geography, construct, modality, and resolution matter? What operation will be performed? What is the cost of error? Which transformations and combinations are proposed?

Against this context, usage evidence can answer practical questions. Who has used the data, and for what research? What findings or products resulted? Which other datasets, software, and models were used with it? How diverse and independent are the user communities? What limitations recur in the literature? Have uses been reproduced, challenged, or corrected? What has changed across versions and over time?

The resulting trust signal should be a profile rather than a universal verdict. It can summarize dimensions such as production accuracy, coverage, reliability, objectivity, reputation, protection, semantic alignment, operational usability, community evidence, and purpose fit. Each dimension should carry its evidence, uncertainty, timestamp, version, and scope. Where evidence is absent, the system should say ``insufficient evidence,'' not silently substitute a neutral or favorable score.

This approach complements established infrastructure. FAIR principles improve findability, accessibility, interoperability, and reuse \cite{wilkinson2016}. Provenance standards such as PROV-DM and PROV-O describe entities, activities, and agents \cite{moreau2013,provO2013}. Data-quality frameworks identify important production properties, and data cards document intended uses and limitations. Usage graphs add revealed practice: empirical evidence of how communities have actually used, combined, and interpreted data.

\section{A Prototype Framework: The National Data Platform and The Data Insights Pipeline}

\subsection{The National Data Platform}

The National Data Platform (NDP)~\footnote{https://nationaldataplatform.org/} provides a concrete frame for implementing this vision \cite{parasharAltintas2023,parashar2026cni}. NDP is conceived as a broad, federated, and extensible data ecosystem built on existing national infrastructure. Its purpose is not to replace repositories or centralize all scientific data. It links distributed resources and provides shared services for discovery, access, integration, analysis, collaboration, validation, and reuse.

The architecture has two complementary elements. The NDP Hub provides a central entry point for authentication, catalog search, conceptual and geospatial discovery, collaborative projects, workspaces, classrooms, and data challenges. A federation of NDP endpoints allows data providers and computing facilities to expose local catalogs and deploy services near data or compute. Endpoints remain under the control of their providers and can connect local data, streaming sources, virtual research environments, distributed data management, remote execution, and third-party services to NDP's core capabilities.

This separation matters for trust. It permits the platform to host and index rich metadata without requiring every source to move to a central store. It also makes governance part of the architecture: access and computation can occur under the policies of the provider, while shared descriptors make resources discoverable and comparable. Collaborative workspaces then connect discovery to action by allowing users to assemble data, software, models, and compute in a reproducible environment.

NDP currently supports diverse data patterns, including repository data, streaming observations, satellite products, environmental sensing, domain-specific workflows, education environments, and customizable community platforms. Across these settings, the same user journey recurs: discover an asset; assess its relevance; bring it into a workspace; combine it with other assets; execute a workflow on suitable infrastructure; and publish or catalog the resulting product. The Data Insights pipeline inserts contextual evidence into that journey at the moment a user is deciding what to trust and use.

\subsection{The Data Insights Pipeline}

The Data Insights pipeline converts traces of scientific practice into data-usage descriptors and a queryable graph using the approach outlined in~\cite{chenarides2026}. Although implementations will vary by corpus and domain, the pipeline has nine logical stages.

\begin{enumerate}
\item \textbf{Seed and scope.} A research community identifies core datasets, aliases, related terms, temporal and geographic boundaries, and a definition of substantive use. Starting with community-named data makes the initial search tractable and grounds it in disciplinary practice.

\item \textbf{Candidate retrieval.} String, identifier, conceptual, and metadata searches retrieve publications or other documents that may reference a seed dataset. This high-recall stage accommodates acronyms, variant names, URLs, and informal references.

\item \textbf{Context filtering.} Domain concepts, dates, languages, affiliations, or other criteria reduce the candidate corpus to the research context of interest. Filtering is essential because the same dataset name or acronym can occur in unrelated fields.

\item \textbf{Mention and use assessment.} A language model examines the full text for direct evidence that the dataset is mentioned and substantively used. The distinction is operational: use requires evidence that the study analyzed, modeled, experimented on, processed, or derived findings from the dataset. Each classification retains an exact supporting excerpt, rationale, confidence, and source location for audit.

\item \textbf{Open discovery.} Within validated publications, the model identifies additional datasets not present in the original seed list. Candidate names are normalized, aliases are consolidated, and domain experts remove generic phrases, survey instruments, or inaccessible objects that do not resolve to distinct sources. The new datasets can seed another search iteration.

\item \textbf{Enrichment.} Publication metadata add authors, affiliations, journals, fields, citations, funders, and time. Further extraction can identify software, models, workflows, variables, joins, transformations, and scientific questions. Persistent identifiers such as DOIs, ORCIDs, RORs, and software identifiers support entity resolution.

\item \textbf{Graph construction and descriptors.} The pipeline represents validated relations as a provenance-bearing graph and aggregates them into dataset-level descriptors. A minimal descriptor can be anchored to a schema.org dataset and include publication-level reviews, mention and use confidence, evidence, and related assets. More expressive implementations can align with Croissant, PROV-O, and domain vocabularies.

\item \textbf{Delivery in NDP.} Search results can signal when contextual insights are available. A dataset page can expose numbers and trends of uses, top publications, user communities, institutions, topics, co-used datasets, and software, with links to the underlying evidence. Related assets become navigable rather than merely listed.

\item \textbf{Feedback and updating.} Users can flag false positives, missing uses, incorrect entity resolution, version errors, or undocumented limitations. Feedback enters a curation queue, and periodic reruns update the graph as the literature and data landscape evolve.
\end{enumerate}

\begin{figure*}[!t]
\centering
\includegraphics[width=0.96\textwidth]{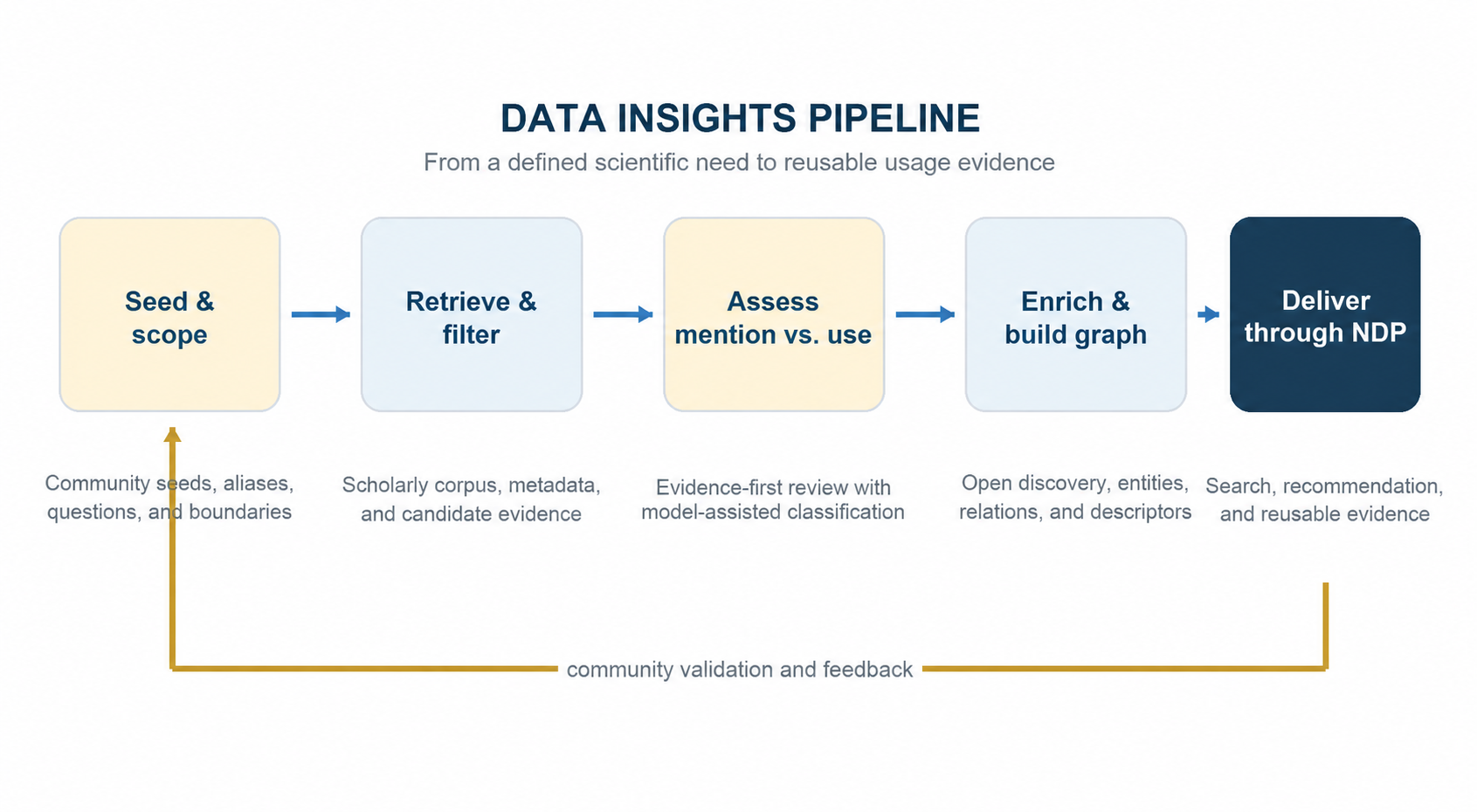}
\caption{The Data Insights pipeline turns community-named data and traces of scientific practice into evidence-bearing descriptors, usage graphs, and discovery services. Language models assist retrieval and extraction, while source evidence and human review preserve auditability.}
\label{fig:pipeline}
\end{figure*}

The pipeline is deliberately evidence-first. Language models assist with retrieval, classification, and extraction; they do not become the source of truth. Off-model evidence, reproducible prompts, model versions, structured outputs, adversarial verification, and human adjudication are required to make the process auditable.

\section{Conclusion}
AI-enabled science makes data discovery both more powerful and more consequential. Systems can search farther than any individual scientist, but distance from familiarity increases the need for context. Catalog metadata, provenance, FAIR practices, and production-quality frameworks remain essential. They should be complemented by evidence about the social life of data: who has used a source, for which purposes, in what combinations, with what tools, and with what results.

Data-usage graphs provide a practical way to organize that evidence. They can illuminate dark data, reveal user communities, document dependencies, identify substitutes, reward data producers, and connect datasets to reusable software and workflows. Implemented through the National Data Platform, the Data Insights data discovery demonstrates how large-scale text analysis, standardized descriptors, federated infrastructure, and community feedback can turn usage traces into discovery and trust services.

The guiding principle is simple: trust should emerge from transparent evidence of production, meaning, use, and stewardship, interpreted for a stated purpose. It should never be reduced to popularity or presented as absolute. A purpose-conditioned trust profile, paired with evidence-aware recommenders and question-to-artifact workflows, can help scientists and AI agents move from ``What data can I find?'' to the more important question: ``What evidence justifies using these digital artifacts to answer this scientific question?''

\section*{Acknowledgment}

This research was funded in part by the US National Science Foundation via awards 2333609, 2440195, and 2504592, and by a subcontract 60070846 on award 2436842. The ideas in this paper have been inspired through conversations with Julia Lane and her writings. The authors used ChatGPT in the development of this manuscript.

\bibliographystyle{IEEEtran}
\bibliography{references}

\end{document}